\pdfoutput=1

\documentclass[prl,reprint,aps,superscriptaddress,nobibnotes,nofootinbib]{revtex4-2}

\usepackage[utf8]{inputenc}
\usepackage{mathptmx}
\usepackage{amsmath}
\usepackage{amssymb}
\usepackage[tight]{units}
\usepackage{color,graphicx}
\usepackage[colorlinks,citecolor=blue,urlcolor=blue,linkcolor=blue]{hyperref}
\usepackage[normalem]{ulem}
\usepackage{enumitem}
\usepackage{mathtools}
\usepackage{braket}
\usepackage{graphicx}
\usepackage{subfig}

\usepackage[dvipsnames]{xcolor}

\begin{document}
\title{Description of CRESST-III lithium aluminate data}
\newcommand{\mpi}{\affiliation{Max-Planck-Institut f\"ur Physik, 80805 M\"unchen, Germany}}
\newcommand{\coimbra}{\affiliation{Also at: LIBPhys, Departamento de Fisica, Universidade de Coimbra, P3004 516 Coimbra, Portugal}}
\newcommand{\hephy}{\affiliation{Institut f\"ur Hochenergiephysik der \"Osterreichischen Akademie der Wissenschaften, 1050 Wien, Austria}}
\newcommand{\ati}{\affiliation{Atominstitut, Technische Universit\"at Wien, 1020 Wien, Austria}}
\newcommand{\tum}{\affiliation{Physik-Department, TUM School of Natural Sciences, Technische Universit\"at M\"unchen, 85747 Garching, Germany}}
\newcommand{\tuebingen}{\affiliation{Eberhard-Karls-Universit\"at T\"ubingen, 72076 T\"ubingen, Germany}} 
\newcommand{\bratislava}{\affiliation{Faculty of Mathematics, Physics and Informatics, Comenius University, 84248 Bratislava, Slovakia}}

\newcommand{\oxford}{\affiliation{Department of Physics, University of Oxford, Oxford OX1 3RH, United Kingdom}}
\newcommand{\wmi}{\affiliation{Also at: Walther-Mei\ss ner-Institut f\"ur Tieftemperaturforschung, 85748 Garching, Germany}}
\newcommand{\lngs}{\affiliation{INFN, Laboratori Nazionali del Gran Sasso, 67010 Assergi, Italy}}
\newcommand{\gssi}{\affiliation{Also at: Gran Sasso Science Institute, 67100, L'Aquila, Italy}}
\newcommand{\cassino}{\affiliation{Also at: Dipartimento di Ingegneria Civile e Meccanica, Università degli Studi di Cassino e del Lazio Meridionale, 03043 Cassino, Italy}}

\newcommand{\saopaulo}{\affiliation{Also at: Instituto de Física da Universidade de São Paulo, São Paulo 05508-090, Brazil}}

\newcommand{\aeth}{\affiliation{Now at: Department of Physics, ETH Zurich, CH-8093 Zurich, Switzerland}}
\newcommand{\apsi}{\affiliation{Now at: ETH Zurich - PSI Quantum Computing Hub, Paul Scherrer Institute, CH-5232 Villigen, Switzerland}}

\mpi
\lngs
\tum
\hephy
\ati
\tuebingen
\oxford

\coimbra
\gssi
\wmi
\cassino

\author{G.~Angloher}
  \mpi

\author{S.~Banik}
  \hephy
  \ati

\author{G.~Benato}
  \gssi 

\author{A.~Bento}
  \mpi
  \coimbra 

\author{A.~Bertolini}
\email[Corresponding author: ]{anbertol@mpp.mpg.de}
  \mpi

\author{R.~Breier}
  \bratislava

\author{C.~Bucci}
  \lngs 

\author{J.~Burkhart}
  \hephy
  \ati

\author{L.~Canonica}
  \mpi 

\author{A.~D'Addabbo}
  \lngs

\author{S.~Di~Lorenzo}
  \lngs

\author{L.~Einfalt}
   \email[Corresponding author: ]{leonie.einfalt@oeaw.ac.at}
  \hephy
  \ati
  
\author{A.~Erb}
  \tum
  \wmi
  
\author{F.~v.~Feilitzsch}
  \tum

\author{N.~Ferreiro~Iachellini}
  \mpi  
  
 \author{S.~Fichtinger}
  \hephy
 
\author{D.~Fuchs}
  \mpi  
 
\author{A.~Fuss}
  \hephy
  \ati

\author{A.~Garai}
  \mpi 
  
 \author{V.M.~Ghete}
  \hephy 

\author{P.~Gorla}
  \lngs 

\author{P.V.~Guillaumon}
  \mpi
  \saopaulo

 \author{S.~Gupta}
 \email[Corresponding author: ]{shubham.gupta@oeaw.ac.at}
  \hephy 

\author{D.~Hauff}
  \mpi 

\author{M.~Ješkovsk\'y}
  \bratislava

\author{J.~Jochum}
  \tuebingen 

\author{M.~Kaznacheeva}
  \tum

\author{A.~Kinast}
  \tum
  
\author{H.~Kluck}
  \hephy
  \ati

\author{S.~Kuckuk}
  \tuebingen 

\author{H.~Kraus}
  \oxford

\author{A.~Langenk\"amper}
  \tum
  \mpi

\author{M.~Mancuso}
  \mpi
 
 \author{L.~Marini}
  \lngs

\author{L.~Meyer}
  \tuebingen 
  
\author{V.~Mokina}
  \hephy
 
\author{A.~Nilima}
  \mpi 

\author{M.~Olmi}
  \lngs
  
\author{T.~Ortmann}
  \tum

\author{C.~Pagliarone}
  \lngs 
  \cassino

\author{L.~Pattavina}
  \tum
  \lngs

\author{F.~Petricca}
  \mpi 

\author{W.~Potzel}
  \tum 

\author{P.~Povinec}
  \bratislava

\author{F.~Pr\"obst}
  \mpi

\author{F.~Pucci}
  \mpi 
  
\author{F.~Reindl}
  \hephy
  \ati

\author{J.~Rothe}
  \tum
  
\author{K.~Sch\"affner}
  \mpi

\author{J.~Schieck}
  \hephy
  \ati 

\author{D.~Schmiedmayer}
   \hephy
   \ati

\author{S.~Sch\"onert}
  \tum 
  
\author{C.~Schwertner}
  \hephy
  \ati

\author{M.~Stahlberg}
  \mpi

\author{L.~Stodolsky}
  \mpi 

\author{C.~Strandhagen}
  \tuebingen

\author{R.~Strauss}
  \tum

\author{I.~Usherov}
  \tuebingen 

\author{F.~Wagner}
  \email[Corresponding author: ]{felix.wagner@phys.ethz.ch}
  \hephy
  \aeth
  \apsi

\author{M.~Willers}
  \tum 

\author{V.~Zema}
  \mpi

\collaboration{CRESST Collaboration}
\noaffiliation

\begin{abstract}
Two detector modules with lithium aluminate targets were operated in the CRESST underground setup between February and June 2021. The data collected in this period was used to set the currently strongest cross-section upper limits on the spin-dependent interaction of dark matter (DM) with protons and neutrons for the mass region between 0.25 and 1.5~GeV/c$^2$. The data are available via Ref.~\cite{DMDC:CRESST}. In this document, we describe how the data set should be used to reproduce our dark matter results.  
\end{abstract}
\maketitle

\section{Introduction}
\label{sec:intro}

The spin-dependent DM results presented in Ref.~\cite{PhysRevD.106.092008} were obtained from the data sets described in this work. For a general description of CRESST-III detectors and the analysis procedure, we refer to Ref.~\cite{cresstcollaboration2020description}. The operated lithium aluminate (LiAlO$_2$) targets had a weight of 10.46 g. Two different modules were operated, named Li1, in which phonon signals from the LiAlO$_2$ crystal and the scintillation light were read out in parallel, and Li2 in which only the phonon signal was read out. The DM data set was taken with an appropriate blinding scheme (as explained in Ref.~\cite{PhysRevD.106.092008}) for a total exposure of 1.161 and 1.184 kg days for Li1 and Li2, respectively. The energy thresholds were determined as (83.60 $\pm$ 0.02) eV and (94.09 $\pm$ 0.13) eV, and the baseline energy resolutions as (13.10 $\pm$ 0.02) eV and (15.89 $\pm$ 0.18) eV, respectively.

For the matrix elements used for the spin-dependent interaction and astrophysical constants, as well as further information on the experimental setup, raw data analysis, and event selection, we refer to ~\cite{PhysRevD.106.092008}. 

We describe the data files that were released in Ref.~\cite{DMDC:CRESST} in Sec.~\ref{sec:files} and compare the limits calculated with these (simplified) released data files with the official limits from Ref.~\cite{PhysRevD.106.092008} in Sec.~\ref{sec:results}. It is recommended to follow the detailed description of our analysis in Ref.~\cite{PhysRevD.106.092008} when using this data for additional studies.

\section{Data files}
\label{sec:files}

\begin{figure}[!t]
\centering
\includegraphics[width=\linewidth]{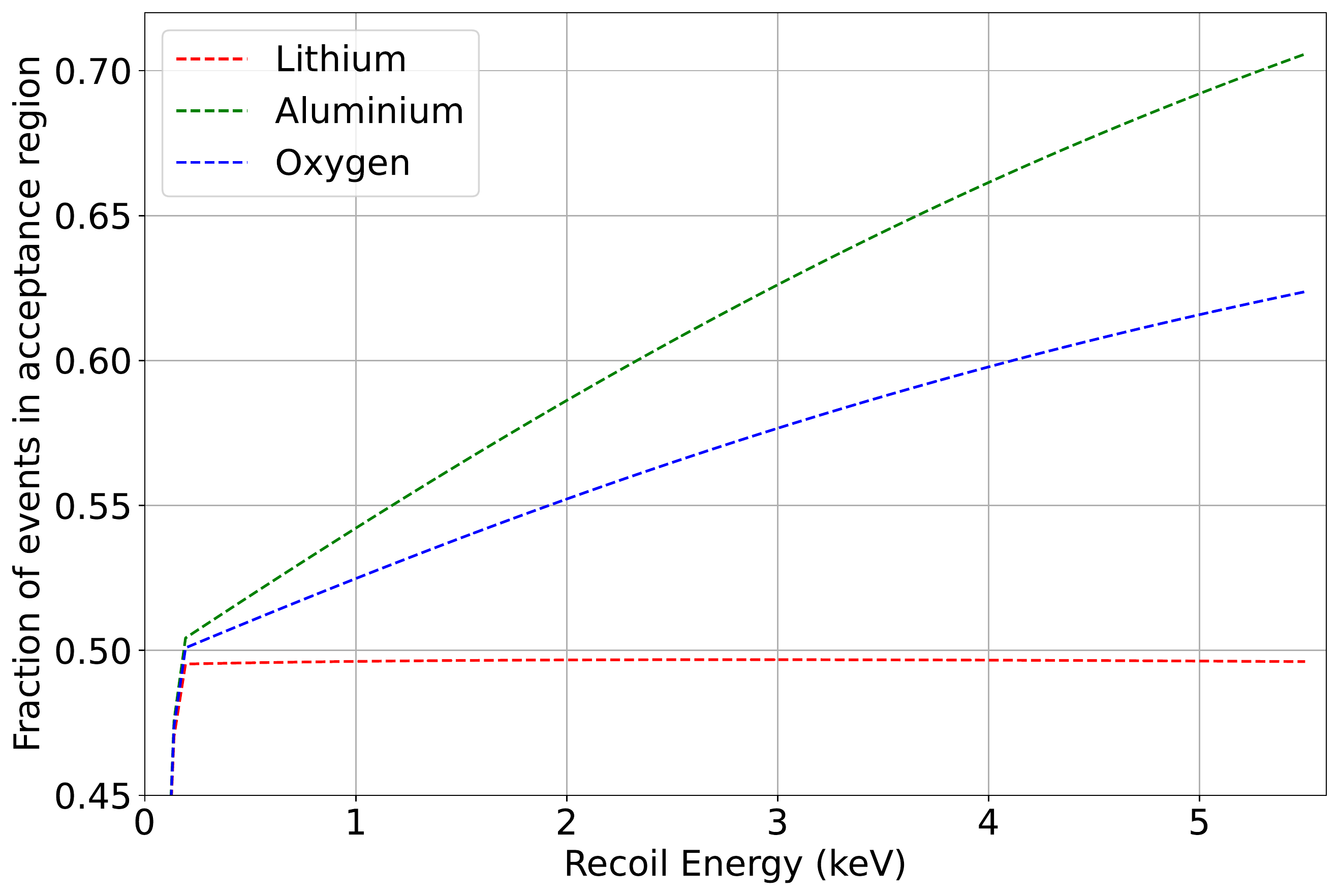}
\caption{Overlap of the nuclear recoil bands with the acceptance region for Li1.}
\label{fig:acceptance_probs}
\end{figure}

The data release contains the following files:

\begin{enumerate}
    \item \label{enum:point1} The files \verb|C3_Li1_Fulldata.xy| and \verb|C3_Li2_Fulldata.xy| contain a list of the recoil energies of all events after data selection cuts for Li1 and Li2, respectively. For Li1, these events correspond to the spectrum shown in Ref.~\cite{PhysRevD.106.092008}, Fig.~7 (left, black). Note that the shown spectrum is corrected with the cut efficiency, while the released data contains the individual event energies and does therefore not yet include the information contained in the cut efficiency.
    \item The file \verb|C3_Li1_AR.txt| contains a list of the recoil energies of the events in the light-yield vs.~energy acceptance region (AR) for Li1. These events correspond to the spectrum shown in Ref.~\cite{PhysRevD.106.092008}, Fig.~7 (left, red) that is also corrected for the cut efficiency. Since there is no operating light channel for Li2, all the events are considered as candidate events to calculate the exclusion limits, and the acceptance region is not defined for this module. 
    \item The files \verb|C3_Li1_cuteff.txt| and \verb|C3_Li2_cuteff.txt| contain the binned cut efficiencies (cuteff) for Li1 and Li2, respectively, which are also plotted in Ref.~\cite{PhysRevD.106.092008}, Fig.~5 (black). 
    \item The acceptance region for the DM searches was defined in Ref.~\cite{PhysRevD.106.092008} as the lower half of the recoil band of the lightest nucleus, to minimize leakage from electromagnetic background. 
    Therefore, the nuclear recoil bands overlap only partly with the acceptance region in the light-yield vs.~energy plane. This energy-dependent overlap for all three nuclei is contained as a binned histogram in the files \verb|C3_Li1_eff_AR_Al.txt|, \verb|C3_Li1_eff_AR_Li.txt| and \verb|C3_Li1_eff_AR_O.txt| for Li1. In this form, it can be included in the limit calculation by multiplying the expected signal of the corresponding nucleus with the cut efficiency and this energy-dependent overlap. The acceptance region is shown in Ref.~\cite{PhysRevD.106.092008}, Fig.~6 (right, green). The content of the files is shown in Fig.~\ref{fig:acceptance_probs}. We do not show this fraction separately for $^6$Li and $^7$Li as their quenching factors are expected to be almost identical and thus have a similar overlap with the acceptance region. 
    \item The official exclusion limits shown in Ref.~\cite{PhysRevD.106.092008}, Fig.~8, are contained in the files \verb|C3_Li1_Limits_SD_Neutron.txt| and \verb|C3_Li1_Limits_SD_Proton.txt| for Li1, and \verb|C3_Li2_Limits_SD_Neutron.txt| and \verb|C3_Li2_Limits_SD_Proton.txt| for Li2.
\end{enumerate}

\noindent

The released data is simplified and compared to the data used in Ref.~\cite{PhysRevD.106.092008}: on the one hand, we do not release the light-yield information explicitly, since it is already sufficiently contained in the overlap of the recoil bands with the acceptance region. On the other hand, we also do not release all simulated events individually since their information is already contained in the cut efficiency. The method applicable for the limit calculation with the released data is, therefore, slightly different from the method used for the exclusion limit in Ref.~\cite{PhysRevD.106.092008}, and we will refer to it as a simplified method. Hence, we expect a slight deviation when calculating the exclusion limits with the reduced data set. The results and methods are compared in the following section.

\section{Comparison with official limits}
\label{sec:results}

\begin{figure*}[t]
\centering
\subfloat[\centering]{\includegraphics[width=\linewidth]{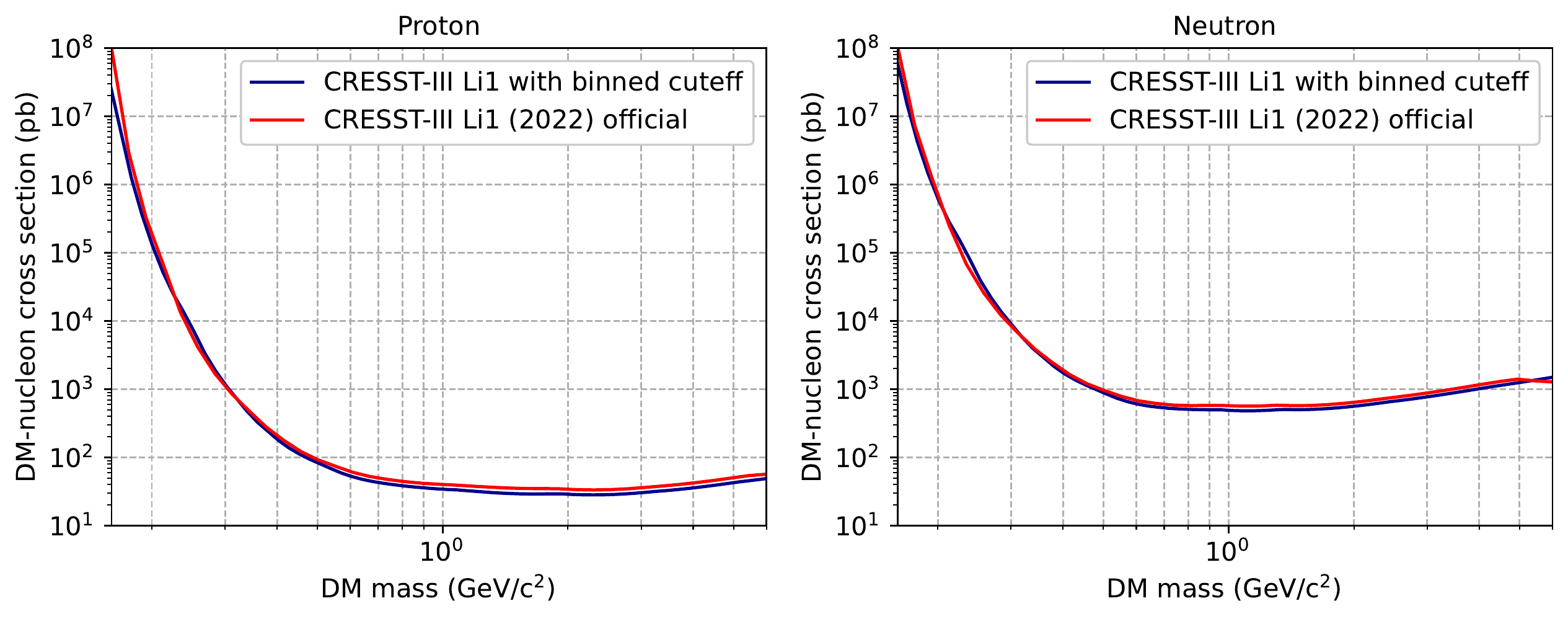}}
\quad
\subfloat[\centering]{\includegraphics[width=\linewidth]{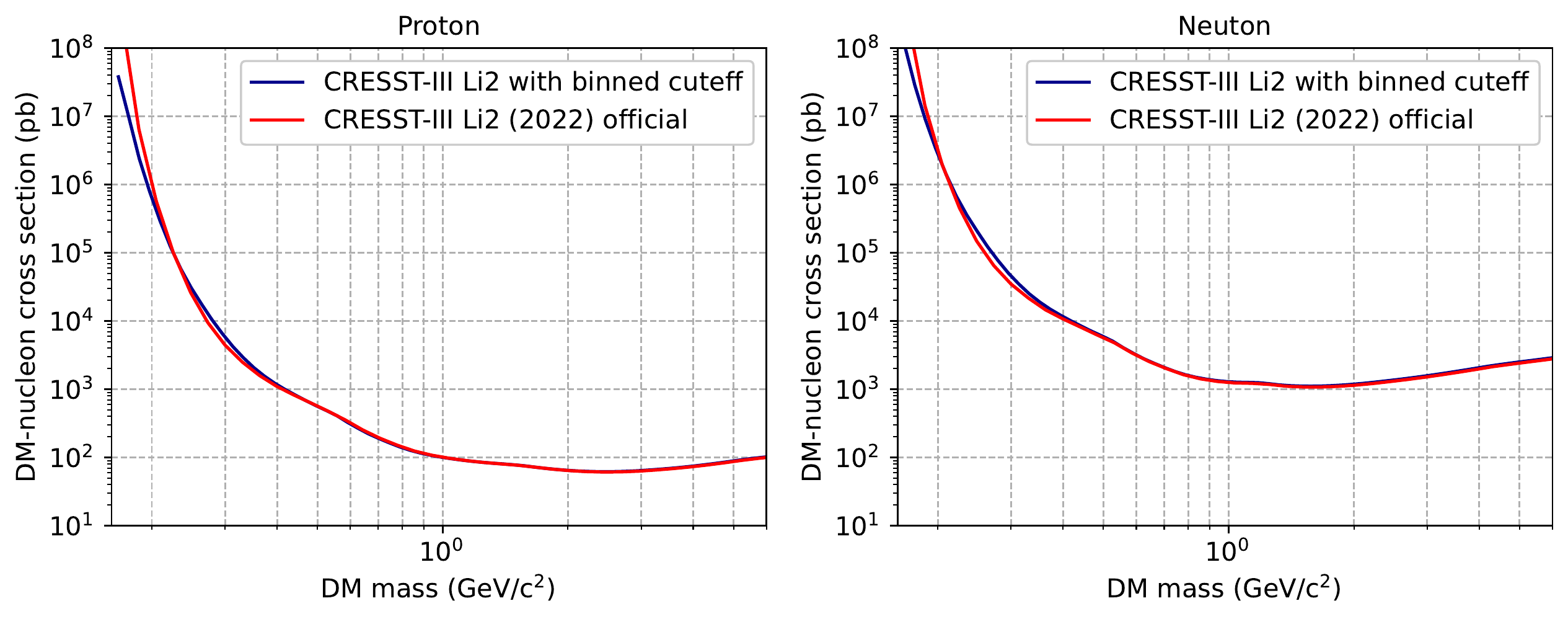}}
\caption{Cross-section upper limits for spin-dependent interactions of DM with protons (left) and neutrons (right) for (a) Li1 and (b) Li2. The official limits from Ref.~\cite{PhysRevD.106.092008} are shown in red, and limits calculated with the binned cut efficiency from this data release and an analytic convolution for the energy resolution are shown in blue. Only a minor deviation at the lowest masses is visible.}
\label{fig:limits}
\end{figure*}

The official exclusion limits from Ref.~\cite{PhysRevD.106.092008} are calculated by exploiting the full information of the efficiency simulation. Events are simulated following a simplified energy spectrum, either flat in energy or with exponentially more events towards lower energies. These simulated events are processed with the same analysis chain as the measured events. 
To avoid repetition of the simulation for each individual DM mass, and reduce the computational cost of the method, we perform the simulation once with a generic spectrum and reweight the simulation results later with the expected spectrum from DM spectrum. 
By doing so, all effects of the detector, especially the finite energy resolution and the analysis chain, are already included in the simulation.

The data released here can be used to obtain almost identical results in a simplified way. The ground truth spectrum expected from DM scattering can be convoluted with an analytical Gaussian function for the baseline energy resolution. Afterwards, a point-wise multiplication with the cut efficiency and the overlap with the acceptance region for the spectra from individual nuclei can be performed. The advantage of this method is that the full simulated data sets and light yield information are not needed. 

A comparison of the limits obtained with the two methods is shown in Fig.~\ref{fig:limits}. The results are in very good agreement, with residuals at the \%-level over most of the region of interest. The remaining deviations occur mainly for the lightest DM masses. The results of an exclusion limit calculation are generally sensitive to the cut efficiency and possibly the optimization procedures used e.g.~in band- and likelihood fits. Depending on these details, individual events at the edge of the acceptance region may be included or excluded, leading to small deviations in the result. The results compared here were obtained with different software implementations, which explains the remaining deviations in the exclusion limits shown. 

\section{Citation}

If you base your work on our data, we kindly ask you to cite this document as well as Ref.~\cite{PhysRevD.106.092008}. 

\section*{Acknowledgements}
We are grateful to LNGS-INFN for their generous support of CRESST. This work has been funded by the Deutsche Forschungsgemeinschaft (DFG, German Research Foundation) under Germany's Excellence Strategy – EXC 2094 – 390783311 and through the Sonderforschungsbereich (Collaborative Research Center) SFB1258 ‘Neutrinos and Dark Matter in Astro- and Particle Physics’, by the BMBF 05A20WO1 and 05A20VTA and by the Austrian Science Fund (FWF): I5420-N, W1252-N27 and FG1 and by the Austrian research promotion agency (FFG), project ML4CPD. The Bratislava group acknowledges a support provided by the Slovak Research and Development Agency (projects APVV-15-0576 and APVV-21-0377). 

\bibliographystyle{h-physrev}
\bibliography{main}
\end{document}